\documentclass{iau} 
\usepackage{graphicx}

\title[Massive stars in advanced evolutionary stages] %% give here short title %%
{Massive stars in advanced evolutionary stages, 
and the progenitor of GW150914}

\author[Wolf-Rainer Hamann \& the Potsdam Group]   %% give here short author list %%
{Wolf-Rainer Hamann, Lidia Oskinova, Helge Todt, Andreas Sander,
Rainer Hainich, Tomer Shenar \and Varsha Ramachandran}

\affiliation{Institut f\"ur Physik und Astronomie, 
Universit\"at Potsdam, Germany}

\pubyear{2017}
\volume{329}  %% insert here IAU Symposium No.
\jname{The lives and death-throes of massive stars}
\editors{J.J. Eldridge, ed.}
\begin{document}

\maketitle

\begin{abstract}
The recent discovery of a gravitational wave from the merging of two 
black holes of about 30 solar masses each 
challenges our incomplete understanding of massive stars
and their evolution. Critical ingredients comprise mass-loss, rotation,
magnetic fields, internal mixing, and mass transfer in close binary systems. 
The imperfect knowledge of these factors implies large uncertainties for
models of stellar populations and their feedback. In this contribution
we summarize our empirical studies of Wolf-Rayet populations at different
metallicities by means of modern non-LTE stellar atmosphere models, and
confront these results with the predictions of stellar evolution
models. At the metallicity of our Galaxy, stellar winds are probably too
strong to leave remnant masses as high as $\sim$30\,M$_\odot$, but
given the still poor agreement between evolutionary tracks and
observation even this conclusion is debatable. At the low metallicity of
the Small Magellanic Cloud, all WN stars which are (at least now) single
are consistent with evolving quasi-homogeneously. O and B-type stars, in
contrast, seem to comply with standard evolutionary models without
strong internal mixing. Close binaries which avoided early merging 
could evolve quasi-homogeneously and lead to close compact remnants
of relatively high masses that merge within a Hubble time. 
\keywords{stars: atmospheres,
stars: early-type,
stars: evolution,
stars: fundamental parameters,
Hertzsprung-Russell diagram,
stars: mass loss,
stars: winds, outflows,
stars: Wolf-Rayet
}
\end{abstract}

\firstsection % if your document starts with a section,
\section{Introduction}

At the 14th of September 2015, the advanced LIGO detectors registered for
the first time a gravitational wave \cite[(Abbott \etal\
2016)]{abbott2016a}.  According to the  analysis of the waveform, this
wave testified the event of two merging black holes (BHs) of
$36^{+5}_{-4}\,M_\odot$ and $29^{+4}_{-4}\,M_\odot$ at a distance of
about 400\,Mpc. The immediate conclusion, and even the prediction prior
to the measurement, was that such heavy BHs can only form by stellar
evolution at low metallicity, where the mass-loss due to stellar winds
is low and hence the stellar remnants can be more massive 
\cite[(Belczynski \etal\ 2016)]{belczynski2016}. Still heavily debated
is whether such BHs form separately in dense clusters and then combine
into a close pair by dynamical interactions, or whether they evolve as
close binaries all the time. Since both scenarios have their problems, a
primordial origin has also been suggested (see Postnov, these
proceedings).       

\section{Galactic Wolf-Rayet stars}
\label{sect:gal-wr}

\begin{figure}[b]
% \vspace*{-2.0 cm}
\begin{center}
 \includegraphics[width=3.4in]{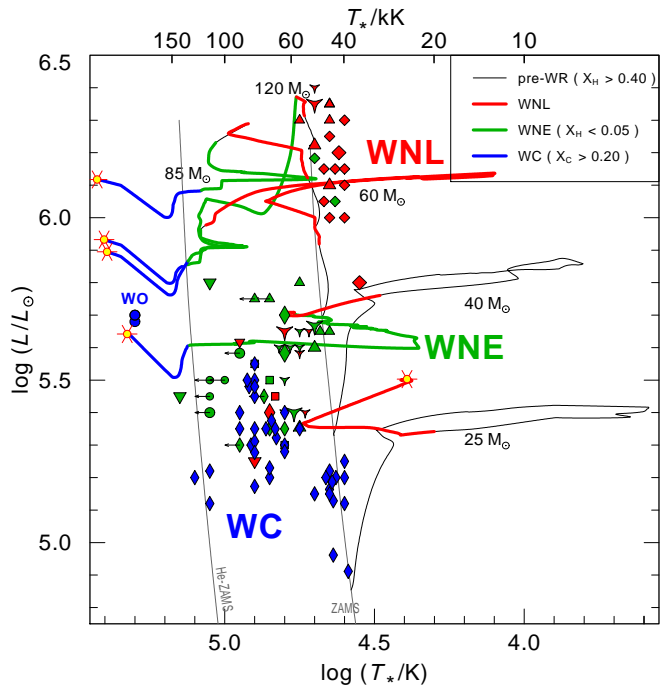} 
% \vspace*{-1.0 cm}
 \caption{HRD of single WR stars in the Galaxy. The discrete symbols
represent analyzed stars \cite[(Hamann \etal\ 2006, Sander \etal\
2012)]{HGL2006}, while the tracks are from the Geneva
group accounting for rotationally induced mixing \cite[(Georgy \etal\
2012)]{georgy2012}.
}
\label{fig:hrd-galwr}
\end{center}
\end{figure}

Massive stars may end their life in a gravitational collapse while being
in the red-supergiant (RGS) phase or as Wolf-Rayet (WR) stars. The
sample of putatively single and optically un-obscured Galactic WR stars 
has been comprehensively analyzed with increasing sophistication (cf.\
Fig.\,\ref{fig:hrd-galwr}). It became clear that the WN stars (i.e.\ the
WR stars of the nitrogen sequence) actually form two distinct groups.
The very luminous WNs with $\log L/L_\odot > 6$ are slightly
cooler than the zero-age main sequence and typically still contain
hydrogen in their atmosphere (often termed WNL for ''late''). In contrast, 
the less luminous WNE stars are hotter (``early'' WN subtypes) and
typically hydrogen free. The WR stars of the carbon sequence (WC) are
composed of helium-burning products and share their location in the
Hertzsprung-Russell diagram (HRD) with the WNE stars. 

From this empirical HRD one can deduce the evolutionary scenario
\cite[(Sander \etal\ 2012)]{sander2012}. The WNL stars evolve directly
from O stars of very high initial mass ($> 40\,M_\odot$). In the mass
range $20 - 40\,M_\odot$ the O stars first become RSGs and then WNE
stars and finally WCs. Stars with initially less then $\approx
20\,M_\odot$ become RSGs and explode there as type II supernova before 
having lost their hydrogen envelope. 

Evolutionary tracks still partly fail to reproduce this empirical HRD
quantitatively, despite of all efforts, e.g.\ with including
rotationally induced mixing. The WNE and WC stars are observed  to be
much cooler than predicted; this is probably due to the effect of
``envelope inflation'' \cite[(Gr\"afener \etal\ 2012)]{graefener2012}.
Moreover, the mass (and luminosity) range of WNE and WC stars is not
covered by the post-RSG tracks. Evolutionary calculations depend
sensitively on the mass-loss rates $\dot{M}$ that are adopted as input
parameters.  Empirical $\dot{M}$ suffer from uncertainties caused by
wind inhomogeneities: when clumping on small scales (``microclumping'')
is taken into account, lower values for $\dot{M}$ are derived from observed
emission-line spectra. Large-scale inhomogeneities (``macroclumping''),
on the other hand, can lead to underestimating mass-loss rates
\cite[(Oskinova \etal\ 2007)]{oskinova2007}. 

Due to the open questions of mixing and the true $\dot{M}$, it is still
uncertain which is the highest BH mass that can be
produced from single-star evolution at Galactic metallicity. For
instance, the luminosities of the two WO stars included in 
Fig.\,\ref{fig:hrd-galwr} would correspond to masses as high as
$\approx 20\,M_\odot$ if they were chemically homogeneous, while the
displayed evolutionary track for initially $40\,M_\odot$ ends with only
$12\,M_\odot$ at core collapse.

%-----------------------------------------------------------------
%-----------------------------------------------------------------

\section{Massive stars at low metallicity}
\label{sect:smc}

\begin{figure}[b]
% \vspace*{-2.0 cm}
\begin{center}
 \includegraphics[width=3.4in]{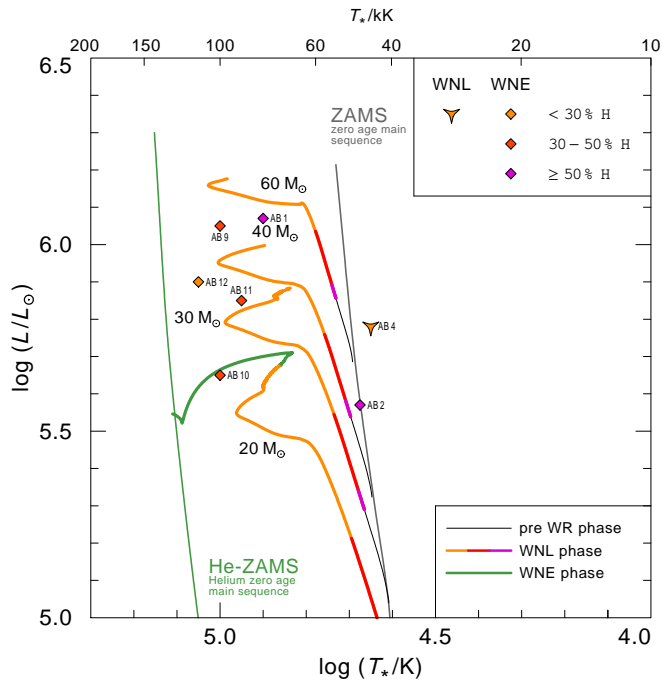} 
% \vspace*{-1.0 cm}
\caption{WN stars in the SMC; the discrete symbols are from the
analyses of the stars as labeled, while the tracks from
\cite{brott2011} adopt very strong rotational mixing (initial $v_{\rm
rot} > 500$\,km/s). Colors code in both cases for the hydrogen mass
fraction (see inlet).  From \cite{hainich2015}
}
\label{fig:hrd-smc-wn}
\end{center}
\end{figure}

The population of massive stars depends critically on their metallicity
$Z$.  This becomes obvious, e.g., from the WR stars in the Small
Magellanic Cloud (SMC) where $Z$ is only about 1/7 of the solar value.
In contrast to the Galaxy, {\em all} putatively single WN
stars in the SMC show a significant fraction of hydrogen in their atmosphere and
wind, like the Galactic WNL stars. However, the WN stars in the SMC
are all hot and compact, located in the HRD
(Fig.\,\ref{fig:hrd-smc-wn}) between the zero age main sequence for
helium stars (He-ZAMS) and the H-ZAMS (or at least, in two cases, 
close the latter). Such parameters cannot be explained with standard
evolutionary tracks, unless very strong internal mixing is assumed which
makes the stars nearly chemically homogeneous. Corresponding tracks 
are included in Fig.\,\ref{fig:hrd-smc-wn}. Quantitatively, they still 
do not reproduce the observed hydrogen mass fractions. 

Stellar winds from hot massive stars are driven by radiation pressure
intercepted by spectral lines. The literally millions of lines from iron
and iron-group elements, located in the extreme UV where the stellar
flux is highest, play a dominant role. Hence a metallicity dependence is
theoretically expected. For O stars, such $Z$ dependence is empirically
established \cite[(e.g.\ Mokiem \etal\ 2007)]{mokiem2007}. For WN stars,
\cite{hainich2015} found a surprisingly steep dependence, probably due
to the multiple-scattering effect (see also Hainich, these proceedings).

%-----------------------------------------------------------------
%-----------------------------------------------------------------

Hence, the lower mass-loss for massive O stars in the SMC, compared to
the Galaxy, might reduce the angular-momentum loss and thus maintain the
rapid rotation which causes the mixing and quasi-homogeneous evolution
to the WR regime. Alternatively, one might speculate that the low
$\dot{M}$ in the SMC is insufficient to remove the hydrogen envelope, 
and thus prevents the formation of single WR stars. This would imply 
that the observed single WNs have all formed through the binary channel, 
possibly as merger products. 

\begin{figure}[b]
% \vspace*{-2.0 cm}
\begin{center}
 \includegraphics[width=3.4in]{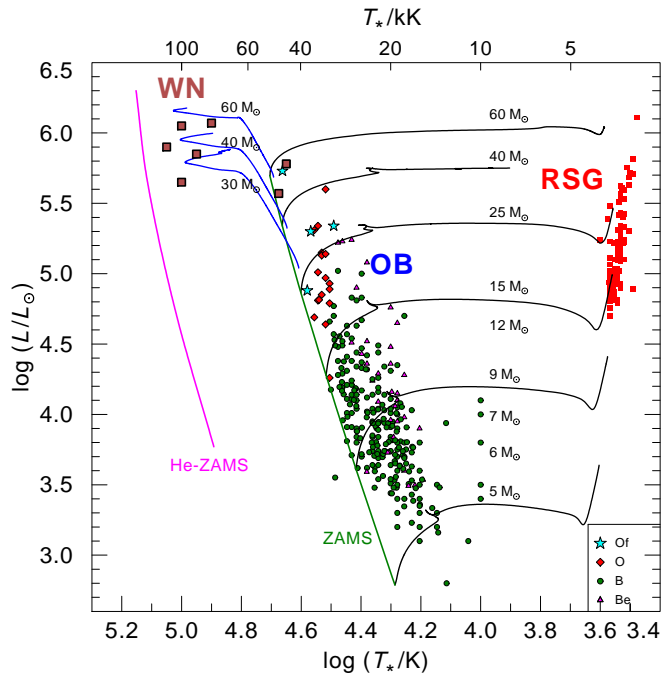} 
% \vspace*{-1.0 cm}
\caption{Massive stars in the SMC. The discrete symbols are from
our analyses of about 300 O and B type stars (Ramachandran \etal\ in
prep.), and the red supergiants are from \cite{massey2003}.  
The WN stars are the same as in Fig.\,\ref{fig:hrd-smc-wn}, as are the
quasi-homogeneous evolutionary tracks. The ``normal'' tracks are also from
\cite{brott2011} but with slower initial rotation ($v_{\rm
rot} = 300$\,km/s).
}
\label{fig:hrd-smc-massivestars}
\end{center}
\end{figure}

But what happens with SMC stars of slightly lower mass? We have analyzed
about 300 OB stars in the region of the supergiant shell SGS\,1
(Ramachandran \etal\ in prep.). Their HRD positions are included in
Fig.\,\ref{fig:hrd-smc-massivestars}, together with ``normal'' tracks
with less rotational mixing. As the comparison shows, the O and
B-type stars with initial masses below $30\,M_\odot$ are consistent with 
``normal'' evolution to the RSG stage.
Only the more massive Of stars might also be consistent with
quasi-homogenous evolution, as are the WN stars discussed above. 

%-----------------------------------------------------------------
%-----------------------------------------------------------------

\section{Massive binaries}
\label{sect:binaries}

In their majority, massive stars are born in binary systems. \cite{marchant2016}
suggested a scenario of ``massive overcontact binary (MOB) evolution'' 
that could lead to a tight pair of massive black holes as observed in
the GW events. Two massive stars which are born as tight binary would
evolve fully mixed due to their tidally induced fast spin and
interaction. They would swap mass several times, making their masses
about equal, but under lucky circumstances they might avoid early
merging. 

\begin{figure}[t]
% \vspace*{-2.0 cm}
\begin{center}
 \includegraphics[width=3.4in]{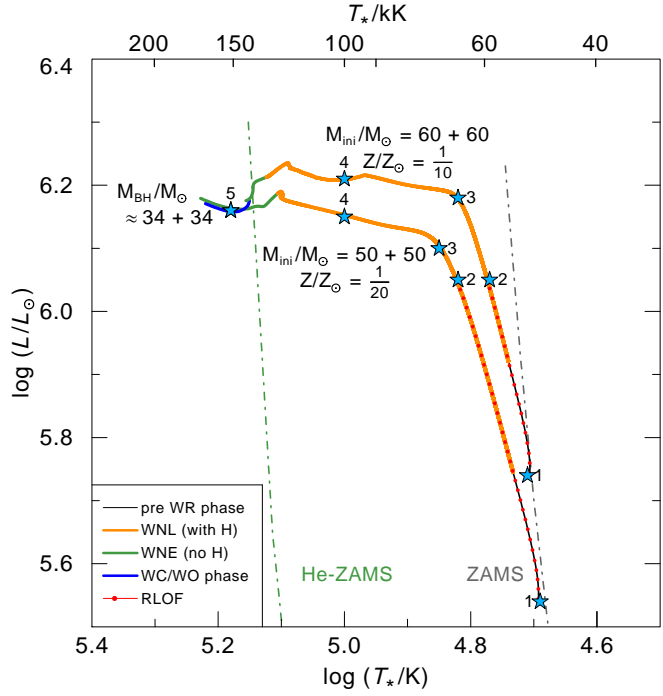} 
% \vspace*{-1.0 cm}
\caption{Tracks for ``massive overcontact binary evolution (MOB)''
(from Marchant, priv.\ comm.). Due the mass exchange, both binary
components have equal masses and therefore evolve identically. One of
the two tracks shown is for a binary of initially $50 + 50\,M_\odot$ and
metallicity of 1/20 solar, and the another one for $60 + 60\,M_\odot$
and metallicity of 1/10 solar. The asterisks mark positions for which
we calculated representative synthetic spectra (Hainich \etal\ in prep.). 
}
\label{fig:hrd-pablo}
\end{center}
\end{figure}

Figure\,\ref{fig:hrd-pablo} shows two such evolutionary tracks from
Marchant (priv.\ comm.).  In both examples, the tracks end at core
collapse with a pair of $34 + 34\,M_\odot$ objects.  We have calculated
synthetic spectra for representative points along the evolutionary
tracks (marked by asterisks in Fig.\,\ref{fig:hrd-pablo}) and found
that such spectra would look unspectacular if observed; in the advanced
stages, the stars would appear as WN-type (with hydrogen, like those in
the SMC discussed above) or, towards the end of the track for
$60\,M_\odot$, as a hot WC type, always with otherwise weak metal lines
due to the low abundances. The only characteristic differences compared 
to single stars would be the doubled luminosity and, if the orbital
inclination is favorite, the radial-velocity variation in the
double-lined spectrum with short period.

%-----------------------------------------------------------------
%-----------------------------------------------------------------

\noindent
{\small {\em Acknowledgements.} We thank Pablo Marchant for providing
the MOB evolutionary tracks. }

\end{document}